\documentclass[twocolumn,aps,prl,floatfix,superscriptaddress]{revtex4-2}
\usepackage[utf8]{inputenc}
\usepackage{graphicx}
\usepackage{multirow}
\usepackage{float}
\usepackage[pdfencoding=auto, psdextra]{hyperref}
\usepackage{color}
\usepackage{amsmath}
\usepackage{physics}
\usepackage{dsfont}

\newcommand{\vect}[1]{{\ensuremath{\boldsymbol{{#1}}}}}
\newcommand{\vk}{\vect{k}}

\newcommand{\A}{\mathcal{A}}
\newcommand{\ham}{\mathcal{H}}

\newcommand{\nn}{\nonumber \\}
\renewcommand{\Im}{\operatorname{Im}}

\begin{document}
\title{Probing quantum geometry with two-dimensional nonlinear optical spectroscopy}

\author{Paul Froese}
\affiliation{Department of Physics, University of Toronto, Toronto, Ontario M5S 1A7, Canada}

\author{Mark R. Hirsbrunner}
\affiliation{Department of Physics, University of Toronto, Toronto, Ontario M5S 1A7, Canada}

\author{Yong Baek Kim}
\affiliation{Department of Physics, University of Toronto, Toronto, Ontario M5S 1A7, Canada}

\begin{abstract}
Recent studies have shown that the nonlinear optical response of crystalline systems is fundamentally a quantum geometric property.
In this work, we propose two-dimensional coherent spectroscopy (2DCS), which measures the nonlinear conductivity as a function of two independent frequencies using two time-delayed light pulses, as a probe of quantum geometry.
We show how the two-frequency second-order nonlinear conductivity, which is naturally measured by 2DCS, decomposes into distinct quantum geometric contributions.
We identify a term arising from the multi-band quantum connection that does not appear in linear response, and show that it can be measured in isolation by considering specific polarizations and enforcing time-reversal symmetry. 
We explore this finding via model calculations for transition metal dichalcogenides and Sr$_2$RuO$_4$.
Through these examples, we demonstrate how 2DCS enables study of the quantum connection, providing a way to compare the quantum geometry of different materials. We also show that one can gain rough momentum-resolved knowledge of the quantum geometry by varying the chemical potential. 
\end{abstract}

\maketitle
{\it Introduction.---}
The quantum geometry of electronic eigenstates plays a vital role in condensed matter systems. The earliest appearance of quantum geometry in condensed matter was in the integer quantum Hall effect~\cite{klitzingNewMethodHighAccuracy1980}, wherein the integral of the Berry curvature gives the Chern number~\cite{thoulessQuantizedHallConductance1982}. It was later appreciated that that Berry curvature has physical consequences itself: it determines the anomalous velocity that enters the semiclassical equations of motion for a Bloch wavepacket~\cite{changBerryPhaseHyperorbits1995,changBerryPhaseHyperorbits1996,sundaramWavepacketDynamicsSlowly1999}. While the Berry curvature is well-known, recent works have revealed the significance of lesser-known geometric quantities, such as the quantum metric and quantum connection. Identifying further geometry-driven phenomena and finding ways to measure quantum geometry are currently of great interest~\cite{tormaEssayWhereCan2023,yuQuantumGeometryQuantum2025,jiangRevealingQuantumGeometry2025}. 

In this work, we propose two-dimensional coherent spectroscopy (2DCS), which measures the nonlinear optical conductivity as a function of two frequencies using two time-delayed pulses of light~\cite{mukamel}, as a probe of multi-band quantum geometry. 2DCS has been studied for both electric and magnetic nonlinear responses of various systems~\cite{luPhysRevLett.118.207204,wanPhysRevLett.122.257401,choiPhysRevLett.124.117205,parameswaranPhysRevLett.125.237601,mahmoodObservationMarginalFermi2021,nandkishorePhysRevResearch.3.013254,freschTwodimensionalElectronicSpectroscopy2023,hartPhysRevB.107.205143,liOpticalMultidimensionalCoherent2023,brenig2024finitetemperatureelectricfield,G_mez_Salvador_2024,katsumiPhysRevLett.132.256903,qiangPhysRevLett.133.126505,Watanabe_2024,zhangPhysRevB.110.104415,zhangTerahertzFieldinducedNonlinear2024,zhangTerahertzfielddrivenMagnonUpconversion2024,barbalasEnergyRelaxationDynamics2025,kaib2025nonlinearspectroscopymagnonbreakdown,yang2025detectionanyonbraidingpumpprobe}. 2DCS provides richer information than conventional single-frequency measurements and is a natural tool for studying the second-order optical conductivity. Here we demonstrate how the second-order conductivity decomposes into distinct multi-band geometric contributions that can be probed by 2DCS. We further show that selecting specific polarizations and enforcing time-reversal symmetry allows one to isolate the contribution from a particular multi-band quantum geometric quantity, the quantum connection. We explore this fact via model calculations, illustrating how 2DCS can reveal differences in the quantum connection between materials using models of transition metal dichalcogenides (TMDs). Finally, using a model for Sr$_2$RuO$_4$, we demonstrate how tuning the chemical potential allows 2DCS to probe different regions of momentum space, providing rough momentum-resolved information about the geometry.

{\it The quantum geometry of optical response.---}
Geometric objects characterize the local structure of a manifold. A metric, for example, defines a notion of distance, and a connection describes parallel transport of tangent vectors and provides a definition of curvature~\cite{nakaharaGeometryTopologyPhysics2003}.  
Quantum geometry broadly refers to the geometric structure of the manifold of eigenstates that diagonalize a parametric Hamiltonian. In crystalline systems, the parameter is the crystal momentum $\vk$ and the Bloch eigenstates, $|\psi_{n\vk}\rangle = e^{i\vk\cdot\hat{\vect{r}}}|u_{n\vk}\rangle$, form the manifold.

The most thoroughly studied quantum geometric quantities are the Berry connection, $\A_n^a(\vk) = i \langle u_{n\vk}|\partial_{k_a} u_{n\vk}\rangle,$ the Berry curvature, $\Omega_{ab}^n,$ which characterizes how the phase of a Bloch state changes as the momentum $\vk$ is varied, and the quantum metric, $g_{ab}^n (\vk),$ which describes the orthogonality, or ``amplitude distance", between states. The quantum metric and Berry curvature are the real and imaginary parts, respectively, of a quantity known as the quantum geometric tensor (QGT)~\cite{provostRiemannianStructureManifolds1980}, defined as
\begin{equation}
    \begin{aligned}
        \label{eq:QGT}
        Q^n_{ab}(\vk) &= \langle \partial_{k_a} u_{n\vk} | \left( \mathds{1} - |u_{n\vk}\rangle \langle u_{n\vk} |\right) | \partial_{k_b} u_{n\vk} \rangle
        .
    \end{aligned}
\end{equation}
The QGT has recently been shown to play a key role in many seemingly disparate areas of condensed matter, ranging from flat-band superconductivity~\cite{peottaSuperfluidityTopologicallyNontrivial2015,hofmannSuperconductivityPseudogapPhase2020,chenGinzburgLandauTheoryFlatBand2024}, capacitance in insulators~\cite{komissarovQuantumGeometricOrigin2024, vermaInstantaneousResponseQuantum2024, Verma_2025}, the fractional quantum anomalous Hall effect~\cite{royBandGeometryFractional2014, leeBandStructureEngineering2017, meraKahlerGeometryChern2021, ledwithVortexabilityUnifyingCriterion2023}, and, the focus of this work, optical responses~\cite{doi:10.1126/sciadv.1501524,sentefprr2019,PhysRevX.10.041041, sentefprb2021,ahnRiemannianGeometryResonant2022, bhallaResonantSecondHarmonicGeneration2022,chaudharyShiftcurrentResponseProbe2022,tai2023quantumgeometriclightmattercouplingcorrelated,kumarBandGeometryInduced2024, avdoshkinMultistateGeometryShift2024,carmichaelProbingQuantumGeometry2025,mitscherlingGaugeinvariantProjectorCalculus2024}

The geometric nature of optical responses arises from the fact that the electric dipole Hamiltonian,
\begin{align}
        \label{eq:hp}
        \ham'(t) = e \vect{r} \cdot \vect{E}(t) 
        ,
\end{align}
couples the perturbing electric field $\vect{E}(t)$ to the electronic position operator, $\vect{r}$. Because the position operator $r^a$ can be represented as the momentum derivative $i\partial_{k_a}$, it provides a basis of tangent vectors on the manifold of Bloch states~\cite{ahnRiemannianGeometryResonant2022}. Specifically, the interband matrix elements of the position operator in the basis of Bloch states~\cite{blount}, equivalent to the non-Abelian Berry connection $\A_{mn}^a = i\langle u_{m\vk}|\partial_{k_a} u_{n\vk} \rangle$, provide the tangent vectors. Geometric quantities like the quantum metric and quantum connection can be constructed from these tangent vectors.

In addition to $g_{ab}^n$, we can construct another metric tensor on the manifold of Bloch states by taking the natural definition of the inner product between the tangent basis vectors. This produces the multi-band generalization of the QGT,
\begin{equation}
    Q_{ab}^{mn}(\vk) = \A_{nm}^a(\vk) \A_{mn}^b(\vk)
    .
\end{equation}
Following Fermi's golden rule, the multi-band QGT is proportional to the optical transition rate between bands, and so naturally appears in the optical conductivity~\cite{ahnRiemannianGeometryResonant2022}.

Another desirable geometric property of the Bloch manifold is a concept of parallel transport of the tangent vectors, which is characterized by the quantum connection~\cite{ahnRiemannianGeometryResonant2022},
\begin{equation}
    \begin{aligned}
        C^{mn}_{a;bc} &= i\A^{b}_{nm}(\vk)\left(\A_{nn}^a(\vk) - \A_{mm}^a(\vk)\right)\A_{mn}^c(\vk)
        \\
        &+
        \A^{b}_{nm}(\vk)\partial_{k_a} \A^c_{mn}(\vk)
        .
    \end{aligned}
\end{equation}
The quantum connection is proportional to the real-space shift of the wavefunction between different bands, and thus contributes to the optical response as well~\cite{sipeSecondorderOpticalResponse2000,doi:10.1126/sciadv.1501524, Cook_2017}. In contrast to the QGT, which is second order in the position operator, the quantum connection is third order in the position operator, and therefore contributes only to the nonlinear optical response. In the next section we derive the precise dependence of the second-order optical conductivity on the QGT and quantum connection.

{\it Geometric decomposition of second-order nonlinear conductivity.---}
To derive the nonlinear optical conductivity, we consider Hamiltonians of the form $\ham_0 + \ham'(t)$, where $\ham_0$ is a single-particle Bloch Hamiltonian and the perturbation $\ham'(t)$ is the electric dipole Hamiltonian given in (\ref{eq:hp}). Working in the interaction picture, the density matrix $\rho$ follows the equation of motion $i \hbar \partial_t \rho(t) = [\ham'(t),\rho(t)]$ with the equilibrium initial condition $[\rho_0]_{nm} = \delta_{nm}f_n$, where $f_n=f(\varepsilon_n)$ is the Fermi-Dirac distribution for band $n$ and we have suppressed the $\vk$ label. The current $j^a(t)$ is given by $j^a(t) = -e\tr\left(v(t)\rho(t)\right)$, where the trace taken over the single-particle Hilbert space, i.e. $\tr(\cdot) = \sum_n \int_k\langle u_{n\vk}| \cdot |u_{n\vk}\rangle$, and $\int_k$ is a shorthand for $\int\frac{d^d\vk}{(2\pi)^d}$. 

The equation of motion is solved order by order by expanding the density matrix in powers of the electric field~\cite{aversaNonlinearOpticalSusceptibilities,sipeSecondorderOpticalResponse2000,venturaGaugeCovariancesNonlinear2017,parkerDiagrammaticApproachNonlinear2019,holderConsequencesTimereversalsymmetryBreaking2020,Passos_2021}. Likewise expanding the induced current in powers of the field, the second-order contribution is 
\begin{equation}
\label{eq:j2}
    \begin{aligned}
        j_2^{a}(t) &= \sum_{bc} \int d\tau_1 d\tau_2 \sigma^{a;bc}(\tau_1,\tau_2) \\
        &\times E^b(t-\tau_2)E^c(t-\tau_1-\tau_2)
        ,
    \end{aligned}
\end{equation}
where the second-order conductivity $\sigma^{a;bc}$ is given by
\begin{equation}
    \begin{aligned}
        \sigma^{a;bc}(\tau_1,\tau_2) &= \frac{e^3}{\hbar^2} \theta(\tau_1)\theta(\tau_2) \\
        &\times\tr\left([[v^a(\tau_1 + \tau_2),r^b(\tau_1)],r^c]\rho_0 \right)
        .
    \end{aligned}
\end{equation}

Evaluating the commutators and Fourier transforming in both time variables, we find that second-order conductivity naturally decomposes into seven distinct terms,
    \begin{equation}
    \label{eq:all_terms}
    \begin{aligned}
    \sigma^{a;bc}(\omega_1,\omega_2) 
    &= 
    \sigma_\mathrm{Drude}^{a;bc}
    +
    \sigma_\mathrm{an}^{a;bc}
    +
    \sigma_\mathrm{DR}^{a;bc} \\
    &+
    \sigma_\mathrm{HOP}^{a;bc}
    +
    \sigma_C^{a;bc}
    +
    \sigma_\mathrm{inj}^{a;bc}
    +
    \sigma_\mathrm{3B}^{a;bc}
    ,
\end{aligned}
\end{equation}
where we have suppressed the frequency dependence for conciseness. The explicit form of each term is given by
\begin{align*}
    \sigma_\mathrm{Drude}^{a;bc}
    &=
    -\frac{e^3}{\hbar^3}g_0^{\omega_1}g_0^{\omega_2}\int_k \sum_n (\partial_{k_c} f_n) (\partial_{k_b} \partial_{k_a} \varepsilon_n) 
    \\
    \sigma_\mathrm{an}^{a;bc}
    &=
    \frac{e^3}{\hbar^3}g_0^{\omega_1}\int_k\sum_{nm}g_{nm}^{\omega_2}(\partial_{k_c} f_{nm})\varepsilon_{nm}Q_{ab}^{mn}
    \\
    \sigma_\mathrm{DR}^{a;bc}
     &= 
     \frac{e^3}{\hbar^3}\int_k \sum_{nm} g_{mn}^{\omega_1}g_{mn}^{\omega_2} \left(\partial_{k_b} f_{nm} \right)\varepsilon_{nm}Q_{ca}^{mn}
    \\
    \sigma_\mathrm{HOP}^{a;bc}
    &=
    \frac{e^3}{\hbar^3}\int_k \sum_{nm}(\partial_{k_b} g_{mn}^{\omega_1})g_{mn}^{\omega_2} f_{nm}\varepsilon_{nm}Q_{ca}^{mn}
    \\
    \sigma_\mathrm{inj}^{a;bc}
    &=
    \frac{e^3}{\hbar^3} g_0^{\omega_2} \int_k \sum_{nm} g_{mn}^{\omega_1} f_{nm} (\partial_{k_a}\varepsilon_{mn}) Q^{mn}_{cb}
    \\
    \sigma_C^{a;bc}
    &=
    \frac{e^3}{\hbar^3}\int_k \sum_{nm} g_{mn}^{\omega_1}g_{mn}^{\omega_2}  f_{nm} \varepsilon_{nm} C_{b;ac}^{nm}
    \\
    \sigma_\mathrm{3B}^{a;bc}
    &=
    \frac{ie^3}{\hbar^3} \int_k \sum_{nm} g_{mn}^{\omega_1} f_{nm} \A^c_{nm} \nn
    &\times \sum_{l\neq m,n}  \left(g_{ml}^{\omega_2} \varepsilon_{ml}\A_{ml}^a\A_{ln}^b
    -
    g_{ln}^{\omega_2}\varepsilon_{ln}\A^b_{ml}\A^a_{ln}\right).
\end{align*}
Here $f_{nm} = f_n(\vk) - f_m(\vk)$, $\varepsilon_{mn} = \varepsilon_m(\vk) - \varepsilon_n(\vk)$, and the poles are given by $g_0^{\omega} = 1/\omega$ and $g_{mn}^\omega = 1/(\omega + \varepsilon_{mn}/\hbar)$. 
We note that the frequencies $\omega_1$ and $\omega_2$, which arise from Fourier transforming $\tau_1$ and $\tau_2$, differ from the frequency convention often seen in the nonlinear optics literature~\cite{BOYD20081}, where the conductivity is a function of the frequencies $\omega_A$ and $\omega_B$, arising from the Fourier transforms of $t_A = \tau_1+\tau_2$ and $t_B = \tau_2$. These conventions are related by the mapping $\omega_1 = \omega_A, \omega_2 = \omega_A +\omega_B$~\cite{Supplement}. The frequencies $\omega_A$ and $\omega_B$ should be understood as driving frequencies, resulting in a measured signal at $\omega_A + \omega_B$, whereas in our convention, $\omega_1$ and $\omega_2$ correspond to the input and measurement frequencies respectively. We incorporate a phenomenological scattering rate by adding a small imaginary part to each frequency: $\omega_1 \to \omega_1 + i\eta, \omega_2 \to \omega_2 + 2i\eta$, which both accounts for disorder and interaction effects and ensures convergence~\cite{kaplanUnifyingSemiclassicsQuantum2023,bradlynSpectralDensitySum2024}. The factor of 2 for $\omega_2$ arises naturally from our frequency convention.

Each contribution is classified according to its geometric interpretation and pole structure, and the names of these terms are chosen to be consistent with previous studies~\cite{bhallaResonantSecondHarmonicGeneration2022,kumarBandGeometryInduced2024,watanabeChiralPhotocurrentParityViolating2021}. The nonlinear Drude term $\sigma_\mathrm{Drude}$ is peaked at $(\omega_1,\omega_2) = (0, 0)$ and only occurs if a Fermi surface is present. The ``anomalous" (an) term is also a Fermi surface term, but is peaked along the $\omega_1 = 0$ axis and depends on the QGT. In the $\omega_2 \to 0$ limit, it corresponds to the current arising from the anomalous velocity induced by the Berry curvature dipole~\cite{PhysRevLett.115.216806, Du_2021}. The ``doubly resonant" (DR) and ``higher order pole" (HOP) contributions also depend on the QGT, but are peaked along the diagonal, $\omega_1 = \omega_2$, and have more complex pole structures. The DR term is a Fermi surface term, while the HOP contribution is a Fermi sea term that arises from differing group velocities between bands, as $\partial_{k_b} g_{mn}^{\omega_1} \propto \partial_{k_b} \varepsilon_{mn}/\hbar = v_{mm} - v_{nn}$. The ``injection current'' term $\sigma_\mathrm{inj}$ also arises from differing group velocities, but is peaked instead on the $\omega_2 = 0$ axis. This term plays an important role in the photovoltaic effect~\cite{sipeSecondorderOpticalResponse2000,parkerDiagrammaticApproachNonlinear2019}.

The remaining contributions arise from higher-order geometric quantities that only appear in the nonlinear response. The first is the $\sigma_C$ term, named for its dependence on the quantum connection $C_{b;ac}^{nm}$, which contributes on the diagonal $\omega_1=\omega_2$. We dub the remaining contribution the ``three-band" (3B) term, as it is present only in systems with at least three bands. This term is proportional to a gauge-invariant sum of triple products of Berry connections, and, while the locations of its peaks depend on the details of the band structure, this term vanishes on the diagonal~\cite{Supplement}. Although the three-band term does not have a known geometric interpretation in general, it simplifies to the quantum torsion on the $\omega_1$ axis, defined as $T^{mn}_{a;cb} = C^{mn}_{a;cb} - C^{mn}_{b;ca}$~\cite{ahnRiemannianGeometryResonant2022,Jankowski_2024}. In this limit, the three-band term is is already well studied as part of the shift current~\cite{doi:10.1126/sciadv.1501524,chaudharyShiftcurrentResponseProbe2022,avdoshkinMultistateGeometryShift2024,Cook_2017}, so we focus on $\sigma_C$ for the remainder of this work.

We can isolate $\sigma_C$ in the 2DCS response by studying the diagonal components of the conductivity tensor, $\sigma^{a;aa},$  and enforcing time-reversal symmetry, in which case all contributions vanish except for $\sigma_C$ and $\sigma_\mathrm{3B}$. By further specifying to the diagonal of the frequency plane, along which $\sigma_\mathrm{3B}$ vanishes~\cite{Supplement}, we can uniquely identify the $\sigma_C$ contribution to the second-order optical conductivity. We further note that the real part of $\sigma_C$ is odd under time-reversal symmetry, so the response actually arises only from the imaginary part of the quantum connection, which we denote as $\tilde{\Gamma}.$
In short, the diagonal components of the second-order conductivity reduce under time-reversal symmetry to the form
\begin{equation}
    \label{eq:TRS_sigma}
    \sigma^{a;aa}(\omega_1,\omega_2) = \frac{ie^3}{\hbar^3} \int_k \sum_{nm} g_{mn}^{\omega_1}g_{mn}^{\omega_2} f_{nm}\varepsilon_{nm} \tilde{\Gamma}^{nm}_{a;aa},
\end{equation}
thus providing a direct probe of the quantum connection.

Next we briefly review how 2DCS probes the second-order optical conductivity before presenting illustrative model calculations exploring the relation between the quantum connection and the second-order conductivity in the limits discussed above.

\begin{figure}[t]
    \centering
    \includegraphics[width=\linewidth]{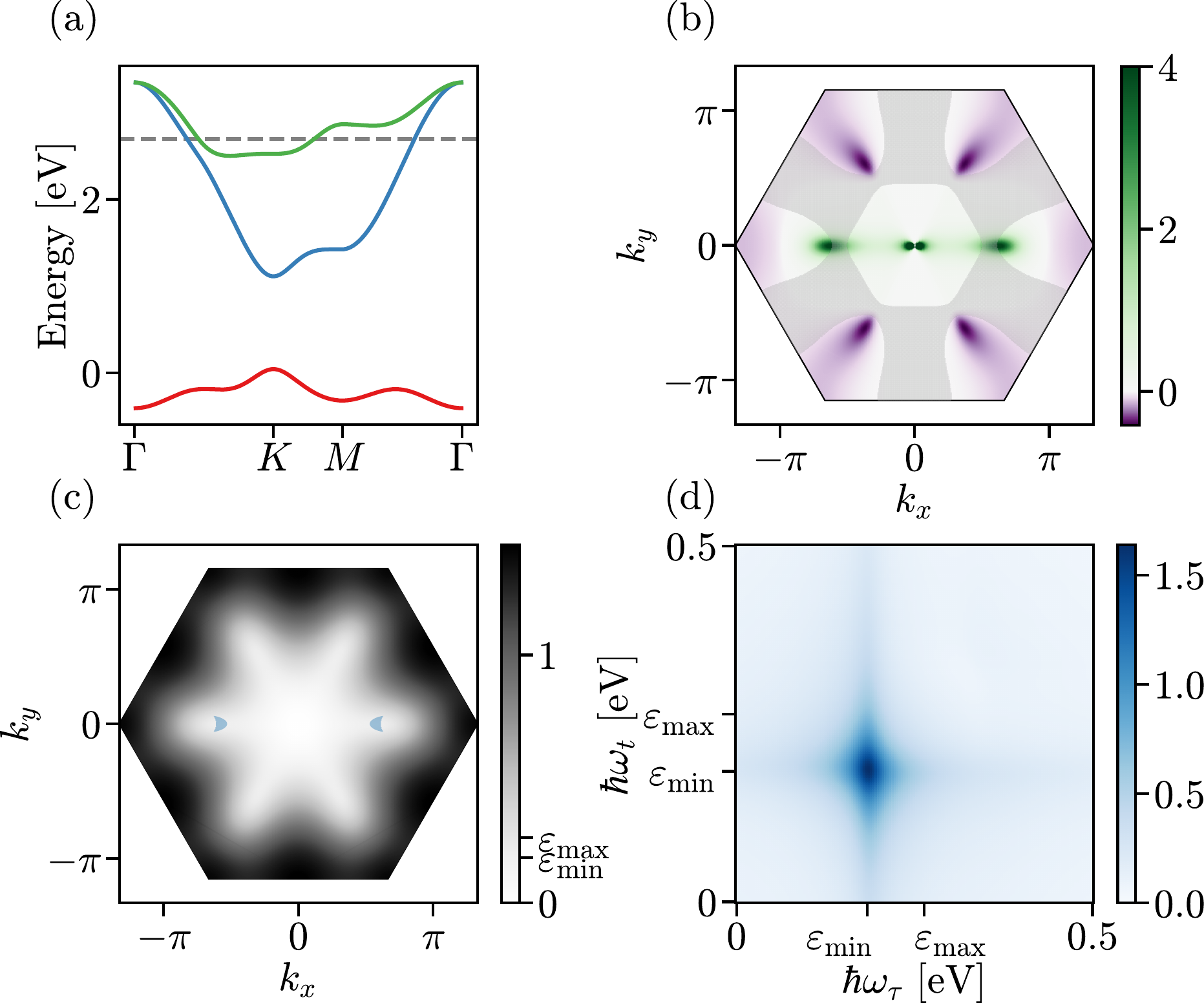}
    \caption{Band structure, quantum connection, and 2DCS response of the NN MoTe$_2$ model.
    (a) The band structure of the model along high symmetry lines in the BZ. The dashed line indicates the chemical potential, $\mu = 2.7~\mathrm{eV}$.
    (b) The imaginary part of the multi-band quantum connection between the upper two bands, $\tilde{\Gamma}^{32}_{y;yy}$. The grey regions correspond to locations where the occupation difference $f_{32}$ is finite.
    (c) The energy difference $\varepsilon_{32}$. The blue shading indicates the region in which $f_{32}\varepsilon_{32}\tilde{\Gamma}^{32}_{y:yy}$ achieves at least $60\%$ of its maximum value. The energies $\varepsilon_\mathrm{min}$ and $\varepsilon_\mathrm{max}$ are the minimum and maximum of $\varepsilon_{32}$ within the highlighted region. (d) The 2DCS response $|\sigma^{y;yy}(\omega_\tau,\omega_t)|$, scaled by factor of $\hbar^3$. In all examples we work at zero temperature, set $e$ and the lattice constant to be 1, and the scattering rate to be $\eta = 0.02~\mathrm{fs}^{-1}$.}
    \label{fig:f1}
\end{figure}

{\it Two-dimensional coherent spectroscopy.---}
In a 2DCS experiment, a sample is illuminated by two strong electric field pulses separated by a delay time $\tau$, and the induced current in the sample is subsequently measured a further time delay $t$ after the second pulse. We refer to the first and second pulses as A and B, respectively. The nonlinear part of the induced current, $j^a_\mathrm{NL}$, can be isolated by subtracting off the response obtained when only the A or B pulse is applied. Taking the ideal limit where the pulses are delta functions, the incoming electric field is given by $ \vect{E}(t) = \vect{E}_\mathrm{A}\delta(t+\tau) + \vect{E}_\mathrm{B}\delta(t)$, and the nonlinear induced current at second-order is given by
\begin{equation}
        j^{a}_{2,\mathrm{NL}} (t) = \sum_{bc} E_A^{c}E_B^{b} \sigma^{a;bc}(\tau,t) 
        .
\end{equation}
A second-order 2DCS experiment is therefore a direct measurement of the second-order conductivity in the time domain. Fourier transforming in both time variables yields the frequency dependent conductivity $\sigma^{a;bc}(\omega_\tau,\omega_t)$, and thus 2DCS probes the second-order response with $(\omega_1,\omega_2) = (\omega_\tau,\omega_t)$.

\begin{figure}
    \centering
    \includegraphics[width=\linewidth]{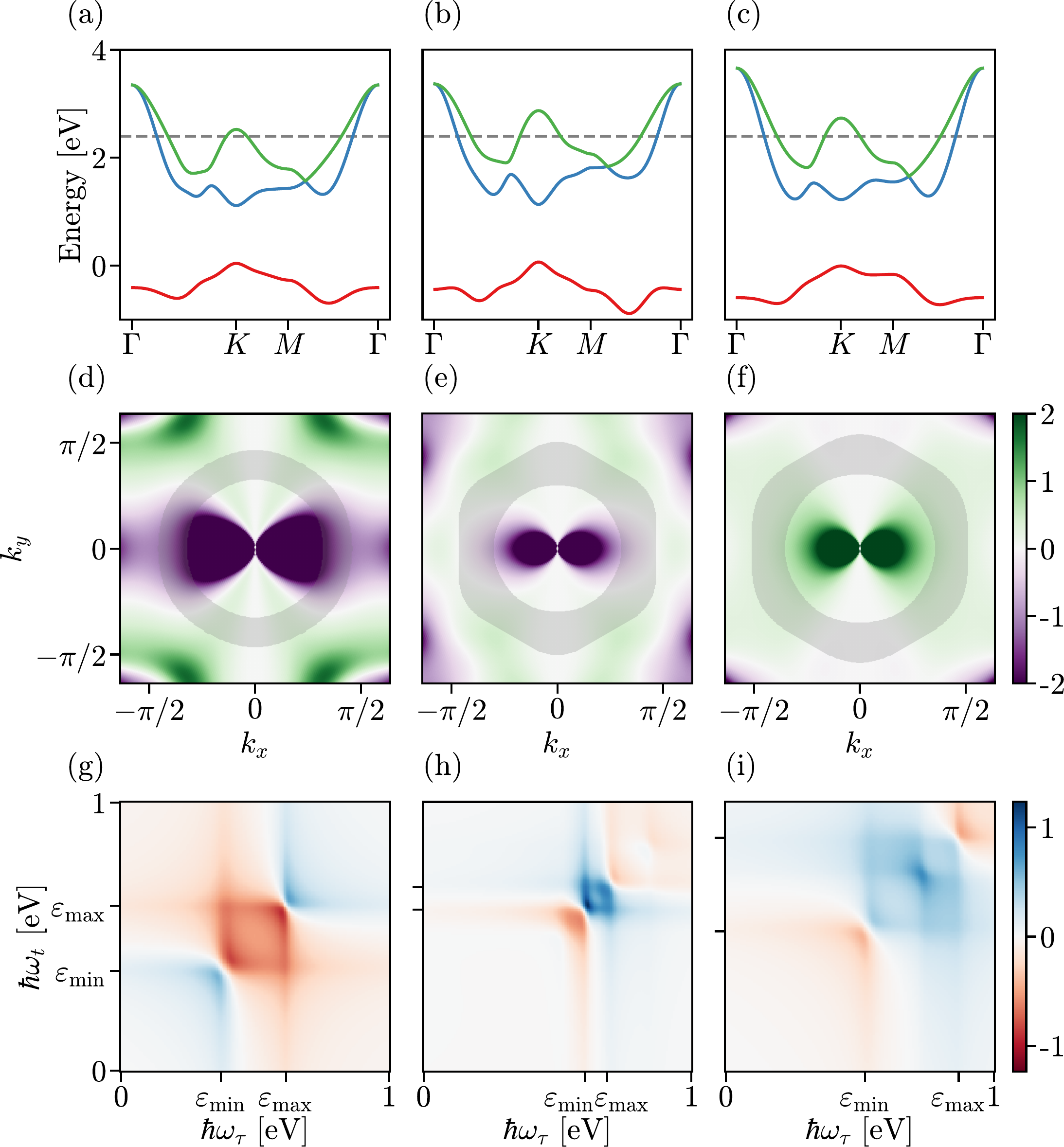}
    \caption{Bandstructure, quantum connection, and 2DCS response for TNN models of MoTe$_2$ and WTe$_2$. The first row depicts the band structures for (a) MoTe$_2$ GGA, (b) WTe$_2$ GGA, and (c) MoTe$_2$ LDA, each with $\mu = 2.4~\mathrm{eV}$ (indicated by the dashed line). (d-f) show $\tilde{\Gamma}^{32}_{y;yy}$ for all models, with the occupation difference $f_{32}$ overlayed, and (g-i) show the imaginary part of the 2DCS response $\Im \sigma^{y:yy}(\omega_\tau,\omega_t)$.}
    \label{fig:f2}
\end{figure}

{\it Transition metal dichalcogenides.---}
TMDs are an ideal platform for studying the quantum geometry of second-order optical response because they break inversion symmetry, a necessary condition for a finite response, and preserve TRS, enabling the isolation of the quantum connection contribution to $\sigma^{a;aa}(\omega_\tau,\omega_t)$~\footnote{The three-band term is also finite, but is numerically small near the diagonal of the frequency where the quantum connection contribution is peaked.}. We utilize symmetry-derived tight-binding models built from the orbitals relevant for the conduction and valence bands of TMDs, with parameters for specific materials obtained by fitting to first-principles (FP) calculations~\cite{liuThreebandTightbindingModel2013}. For each model, we set the chemical potential such that the upper two bands are partially occupied, and focus on the 2DCS response at frequencies corresponding to transitions between the upper bands~\footnote{The chemical potential can be adjusted in a monolayer TMD device via dual-gating, and transitions involving the lowest band are well-separated in frequency from the transitions of interest.}.

We begin by considering MoTe$_2$, retaining only nearest-neighbor (NN) hopping terms in the model. The parameters are obtained from fitting the bandstructure to FP calculations employing the generalized-gradient approximation (GGA). This model is quantitatively accurate only near the $K$ point, but it provides an illustrative example of how the quantum connection can produce a distinct 2DCS response. We plot in Fig.~\ref{fig:f1}~(b) the imaginary part of the quantum connection, $\tilde{\Gamma}^{32}_{y;yy}$, finding that it is strongly peaked along the $\Gamma-K$ line. We also shade in grey the region in which the occupation difference between the upper two bands, $f_{32}(\vk)$, is finite. This is the region that contributes to the 2DCS response, indicating that only the outer two peaks of the quantum connection contribute.

\begin{figure*}
    \centering
    \includegraphics[width=\linewidth]{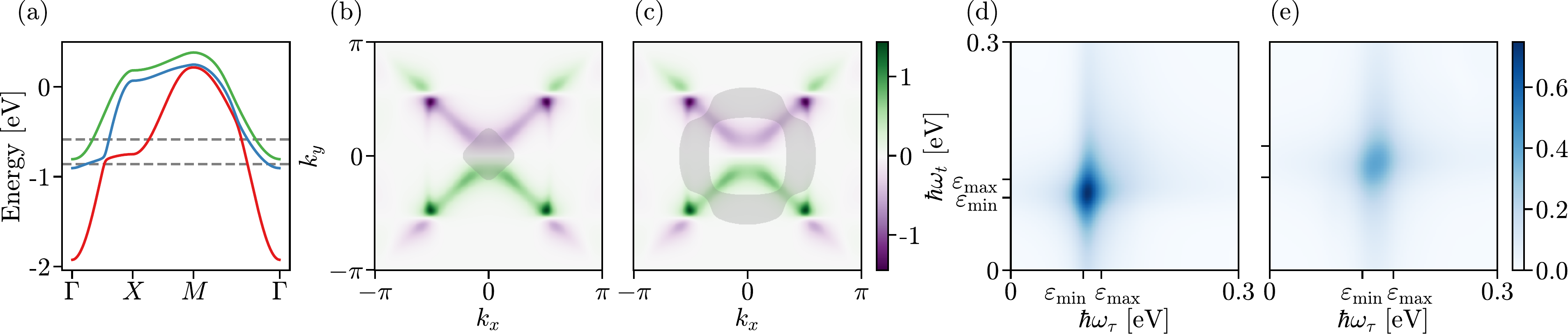}
    \caption{Band structure, quantum connection, and 2DCS response of Sr$_2$RuO$_4$. (a) The band structure along high symmetry lines, with the chemical potentials $\mu_1 = -0.860~\mathrm{eV}$ and $\mu_2 = -0.586~\mathrm{eV}$ marked by the dashed lines. (b, c) The quantum connection $\tilde{\Gamma}^{32}_{x;xx}$ occupation difference $f_{32}$ (gray shading) for $\mu_1$ and $\mu_2,$ respectively. (d, e) The 2DCS response $|\sigma^{x;xx}(\omega_\tau,\omega_t)|$ for $\mu_1$ and $\mu_2$, respectively. The predicted frequency ranges, $\varepsilon_\mathrm{min}$ and $\varepsilon_\mathrm{max}$, are calculated analogously to the previous examples, but now considering the region where $f_{32}\varepsilon_{32}\tilde{\Gamma}^{32}_{x;xx}$ achieves at least $90\%$ of its maximum value.}
    \label{fig:f3}
\end{figure*}

The frequency at which this peak in the quantum connection generates a 2DCS response corresponds to the energy gap between the bands, $\varepsilon_{32}(\vk)$, in the region around the peak. We predict this frequency by calculating the range of the energy gap in the region for which $f_{32}\varepsilon_{32}\tilde{\Gamma}^{32}_{y;yy}$ is above some percentage of its maximum value, heuristically chosen to be 60\%. We plot $\varepsilon_{32}(\vk)$ in Fig.~\ref{fig:f1}~(c), denoting on the colourbar the maximum and minimum values of $\varepsilon_{32}(\vk)$ in the above described region (which is shaded in blue). Finally, we plot $\sigma^{y;yy}(\omega_\tau,\omega_t)$, the 2DCS signal, in Fig.~\ref{fig:f1}~(d). The response exhibits a sharp peak precisely within the frequency range predicted by our above analysis, demonstrating how the quantum connection can lead directly to a strong, predictable response in 2DCS. 

Having established the relation between the quantum connection and the 2DCS response, we proceed to study more realistic TMD models. We consider here tight-binding models including up to third nearest-neighbor (TNN) hopping terms, the bandstructures of which quantitatively agree with FP calculations across the entire Brillouin Zone (BZ)~\cite{liuThreebandTightbindingModel2013}. We begin by comparing the TMDs MoTe$_2$ and WTe$_2$, which have similar bandstructures but qualitatively different quantum geometry. We consider models obtained by fitting to FP calculations using GGA, the bands of which are shown in Fig~\ref{fig:f2}(a)-(b). As above, we tune the chemical potential to isolate transitions between the upper two bands.

We plot the quantum connections of the two models around the $\Gamma$ point in Fig-~\ref{fig:f2}~(d)-(e), revealing differing sign structures. Both MoTe$_2$ and WTe$_2$ exhibit large negative lobes of $\tilde{\Gamma}_{y;yy}^{32}$ near the $\Gamma$ point, extending towards the $K$ point. However, in WTe$_2$ the sign of $\tilde{\Gamma}_{y;yy}^{32}$ changes from negative to positive at larger momenta. As a result, the sign of the quantum connection in the region probed by 2DCS is of opposite sign for the two materials. The sign difference is reflected in the 2DCS response, plotted in Fig.~\ref{fig:f2}~(g)-(h), where $\operatorname{Im} \sigma^{y;yy}$ is negative for MoTe$_2$, and positive for WTe$_2$. We see that in systems where the quantum connection produces peaks in the second-order optical conductivity, 2DCS is capable of measuring the sign of the quantum connection.

In a similar manner, 2DCS can also be used to study the accuracy of different models of a single material. In Fig.~\ref{fig:f2} (c, f, i) we plot the band structure, $\tilde{\Gamma}_{y;yy}^{32}$, and $\operatorname{Im} \sigma^{y;yy}$ of a different model for MoTe$_2$ obtained by fitting to FP calculations employing the local-density approximation (LDA). Interestingly, $\tilde{\Gamma}_{y;yy}^{32}$ for the MoTe$_2$ LDA model has positive lobes emanating from the $\Gamma$ point -- a sign structure opposite to the GGA case. The difference is again clearly revealed in the 2DCS response, with $\operatorname{Im} \sigma^{y;yy}$ positive in the expected energy range for MoTe$_2$ LDA. Measuring the 2DCS response can therefore assess the accuracy of the wavefunctions generated by different models, providing a further metric for the quality of models beyond just the bandstructure.

{\it Inversion symmetry broken strontium ruthenate.---} 
As discussed above, only the part of the BZ with a finite occupation difference between bands contributes to the 2DCS response. Here we explore how one can leverage varying the chemical potential to modify which portion of the BZ participates in the response, thereby providing momentum-resolved access to the quantum connection. To that end, we consider a three-band tight-binding model for Sr$_2$RuO$_4$~\cite{puetterIdentifyingSpintripletPairing2012,Lindquist_2020}, which has a simple band structure, can be doped to tune the chemical potential, and has been successfully studied via 2DCS in a previous experiment~\cite{barbalasEnergyRelaxationDynamics2025}. We break inversion symmetry in the model by hand, which can be accomplished in practice by a suitable choice of substrate~\cite{Supplement}.

We again study transitions between the upper two bands of the model, but consider two different chemical potentials, depicted with dashed lines in Fig.~\ref{fig:f3}~(a). With the lower chemical potential, the region of the BZ which contributes to the response is a small pocket at the $\Gamma$ point, while the higher chemical potential instead probes a ring section of the BZ at larger momenta. We plot $\tilde{\Gamma}^{32}_{x;xx}$ for each chemical potential in Fig.~\ref{fig:f3}~(b, c), with the occupation difference $f_{32}$ overlaid in gray. Because inversion symmetry is only weakly broken, $\tilde{\Gamma}_{x;xx}^{32}$ is nearly completely odd in momentum~\cite{Supplement}. However, the contributions from the positive and negative regions do not entirely cancel out, leading to the finite responses plotted in Fig.~\ref{fig:f3}~(d, e). The peaks in the 2DCS response $|\sigma^{x;xx}(\omega_\tau,\omega_t)|$ appear at different frequencies and have different magnitudes for each chemical potential. This reflects that the energy gap, which determines the frequency of the response, and $\tilde{\Gamma}$, which determines the magnitude, take different values in the two regions of the BZ probed by the different chemical potentials. Therefore, varying the chemical potential allows one to study the relative strength of the quantum connection in different regions of the BZ.

{\it Discussion.---}
In this work, we have addressed two important challenges in the field of quantum geometry: identifying physical phenomena that depend on quantum geometry and devising schemes to directly measure geometric quantities.
To address the first challenge, we clarified how the second-order optical conductivity depends on the quantum geometry, decomposing it into distinct geometric contributions. We utilized this decomposition to address the second challenge by identifying conditions under which the second-order conductivity arises solely from the quantum connection, providing a means to measure it. We illustrated this with model calculations, showing how 2DCS can probe the sign structure and momentum dependence of the quantum connection. 

These examples illustrated how $\sigma_C$ can be isolated in the 2DCS response. However, the generic 2DCS response contains additional geometric information in the remaining terms of (\ref{eq:all_terms}) that we made to vanish by choosing specific polarizations and enforcing time-reversal symmetry. Because these terms lie along different axes in the two-dimensional frequency space, it is possible to distinguish these distinct geometric contributions without forcing the rest to vanish. As such, 2DCS allows one to measure multiple geometric aspects of the conductivity with a single experiment.

Some important future work remains to be done. We considered non-interacting systems, but understanding the role of quantum geometry in the nonlinear conductivity of interacting systems is an important next step. Constructing a geometric decomposition of the third-order conductivity is another natural future step, as the third-order response does not require inversion symmetry breaking, making it relevant to vastly more materials. 

{\it Acknowledgements.--}
We thank Junyeong Ahn for helpful discussions on complex Riemannian geometry and Daniel Schultz for further discussions. We thank Yuan Wan, Peter Armitage, and Kota Katsumi for their helpful comments. This work is supported by the Natural Sciences and Engineering Research Council of Canada (NSERC) and the Centre of Quantum Materials at the University of Toronto. Computations were performed on the Cedar cluster, which the Digital Research Alliance of Canada hosts.

\bibliography{QG_2DCS}

\setcounter{figure}{0}
\let\oldthefigure\thefigure
\renewcommand{\thefigure}{S\oldthefigure}

\setcounter{table}{0}
\renewcommand{\thetable}{S\arabic{table}}

\newpage
\clearpage

\begin{appendix}
\onecolumngrid
	\begin{center}\textbf{\large --- Supplementary Material ---}
    \end{center}

\section{Second-order optical conductivity in the length gauge}
Following the approach of Aversa and Sipe~\cite{aversaNonlinearOpticalSusceptibilities}, here we derive the nonlinear optical conductivity in the length gauge. Further details regarding this derivation can be found in several recent clarifying works~\cite{venturaGaugeCovariancesNonlinear2017,parkerDiagrammaticApproachNonlinear2019}. As introduced in the main text, the system is described by the Hamiltonian $\ham(t) = \ham_0 + \ham'(t)$, where $\ham_0$ is the Hamiltonian of the unperturbed electronic system and the perturbing electric field is included through the electric dipole term $\ham'(t) = e\vect{r}\cdot \vect{E}(t)$. The state of the system is captured by the density matrix $\rho = \sum_\nu p_\nu |\phi_\nu\rangle \langle \phi_\nu |$, where $\nu$ labels the eigenstates of $\ham$ and $p_\nu$ is the probability that the system is in the state $\nu$. We assume that the system is initially described by the equilibrium density matrix $\rho_0$ in the limit $t \to -\infty$. In the interaction picture, the density matrix evolves according to
\begin{align*}
    i\hbar \partial_t \rho(t) = [\ham'(t),\rho(t)]
    .
\end{align*}
Other operators in the interaction picture time-evolve according to $S(t) = e^{i\ham_0 t/\hbar}Se^{-i\ham_0 t/\hbar}$. In this work, we consider non-interacting systems, allowing us to work in a single particle picture.

We solve for optical conductivity by writing down the induced current, $j^a(t) = -e\tr\left(v^a(t) \rho(t)\right)$. To determine the $n$-th order conductivity, we expand both the current and density matrix in powers of the field according to $j^a(t) = \sum_i j_i^{a}(t)$ and $\rho(t) = \sum_i \rho_i(t)$, where $j_i^{a}$ and $\rho_i$ are $i$-th order in $\vect{E}$. By matching powers of the field, we find $j_i^{a}(t) = -e\tr\left(v^a(t) \rho_i(t)\right)$.   

The equation of motion for the density matrix can be solved order by order, yielding:
\begin{align*}
    \rho_0(t) &= \rho_0 \\
    \rho_1(t)  &= \frac{-ie}{\hbar}\sum_a\int_{-\infty}^t dt_1 [r^a(t_1),\rho_0]E^a(t_1) \\
     \rho_2(t) &=
     \frac{-e^2}{\hbar^2}\sum_{ab}\int_{-\infty}^t dt_2 \int_{-\infty}^{t_2} dt_1 [r^a(t_2),[r^b(t_1),\rho_0]]E^a(t_2)E^b(t_1)
     .
\end{align*}
The first- and second-order currents are consequently given by
\begin{align*}
    j_1^{a}(t) &= \frac{ie^2}{\hbar}\sum_b\int_{-\infty}^tdt_1 \tr\left(v^a(t)[r^b(t_1),\rho_0]\right)E^b(t_1) \\
    j_2^{a}(t) &= \frac{e^3}{\hbar^2} \sum_{bc}\int_{-\infty}^t dt_2 \int_{-\infty}^{t_2} dt_1 \tr\left(v^a(t) [r^b(t_2),[r^c(t_1),\rho_0]] \right)E^b(t_2) E^c(t_1)
\end{align*}
We proceed by making a change of variables, introducing new time arguments corresponding to time differences: $\tau_m = t_{m+1} - t_m$, with $t_n :=t$ for the $n$-th order expression:
\begin{align*}
    j_1^{a}(t) &= \frac{ie^2}{\hbar}\sum_b\int_{0}^\infty d\tau_1 \tr\left([v^a(t),r^b(t-\tau_1)]\rho_0\right)E^b(t-\tau_1) \\
    j_2^{a}(t) &= \frac{e^3}{\hbar^2} \sum_{bc} \int_{0}^\infty d\tau_2 \int_{0}^{\infty} d\tau_1 \tr\left([v^a(t),r^b(t-\tau_2)],r^c(t-\tau_1-\tau_2)]\rho_0 \right)E^b(t-\tau_2) E^c(t-\tau_1-\tau_2)
    .
\end{align*}
The first- and second-order optical conductivities are defined according to
\begin{align*}
    j_1^{a}(t) &= \sum_b \int d\tau_1 \sigma^{a;b}(\tau_1)E^b(t-\tau_1) \\
    j_2^{a}(t) &= \sum_{bc} \int d\tau_1 d\tau_2 \sigma^{a;bc}(\tau_1,\tau_2)E^b(t-\tau_2)E^c(t-\tau_1-\tau_2)
\end{align*}
which allows one to read off
\begin{align*}
    \sigma^{a;b}(\tau_1) &= \frac{ie^2}{\hbar}\theta(\tau_1)\tr\left([v^a(\tau_1),r^b]\rho_0\right) \\
    \sigma^{a;bc}(\tau_1,\tau_2) &= \frac{e^3}{\hbar^2}\theta(\tau_1)\theta(\tau_2) \tr\left([v^a(\tau_1 + \tau_2),r^b(\tau_1)],r^c]\rho_0 \right)
\end{align*}
where we have used time-translation invariance to shift the time arguments in the trace.

We now evaluate the commutators appearing in the response functions. First we need the matrix elements of the position operator in the Bloch basis~\cite{venturaGaugeCovariancesNonlinear2017,blount},
\begin{align*}
    \langle \psi_{n\vk}|r^a|\psi_{m \vk'}\rangle = (2\pi)^d \left[-i\delta_{nm}\partial_{k'_a}\delta(\vk'-\vk) + \A^a_{nm}(\vk)\delta(\vk'-\vk)\right]
\end{align*}
The singular intraband part of the matrix element presents a challenge for performing explicit calculations. However, we can circumvent this issue by instead considering commutators between the position operator and $\vk$ - diagonal operators of the form $\langle \psi_{n\vk}|S|\psi_{m\vk'}\rangle = (2\pi)^d\delta(\vk'-\vk) S_{nm}(\vk)$. Such operators are well-behaved and take the form
\begin{align*}
    \langle \psi_{n\vk} | [r^a,S]|\psi_{m\vk'}\rangle = (2\pi)^d\delta(\vk'-\vk) [r^a,S(\vk)]_{nm}
\end{align*}
where we have defined the commutator matrix elements as
\begin{align}
\label{eq:comm}
    [r^a,S(\vk)]_{nm} = i\partial_{k_a}S_{nm}(\vk) + [\A^a(\vk),S(\vk)]_{nm}
    .
\end{align}
Time evolving the position operator in the interaction picture yields the related identity~\cite{parkerDiagrammaticApproachNonlinear2019}
\begin{align*}
    [r^a(\tau),S(\vk)]_{nm} &= (i\partial_{k_a} +\tau \partial_{k_a}\varepsilon_{nm}/{\hbar}) S_{nm}(\vk) \\
    &+ \sum_l \left[
    e^{i\varepsilon_{nl}(\vk)\tau/\hbar}\A^a_{nl}(\vk)S_{lm}(\vk) - S_{nl}(\vk)\A^a_{lm}(\vk)e^{i\varepsilon_{lm}(\vk)\tau/\hbar} \right]
\end{align*}
From now on we suppress momentum labels to keep notation compact. The identity (\ref{eq:comm}) and be used to calculate the matrix elements of the velocity operator, since generally $v^a := \frac{i}{\hbar}[\ham_0,r^a]$, yielding 
\begin{align*}
    v_{nm}^a = \frac{1}{\hbar} \left(\delta_{nm}\partial_{k_a}\varepsilon_n + i\varepsilon_{nm}\A^a_{nm}\right)
\end{align*}
We also use the above commutator identities to evaluate the nested commutators appearing in the conductivities. In the single particle picture, the equilibrium density matrix is given by $[\rho_0]_{nm} = \delta_{nm} f_n$, where $f_n$ is the Fermi distribution function. Meanwhile, the trace is taken over the single particle Hilbert space, i.e. $\tr( \cdot) \to \int_k \sum_n \langle u_{n\vk} | \cdot | u_{n\vk} \rangle$. After Fourier transforming the time arguments, the linear optical conductivity in the frequency domain becomes
\begin{align*}
    \sigma^{a;b}(\omega_1) = \frac{ie^2}{\hbar^2}\int_k \bigg\{\sum_n f_n \frac{\partial_{k_b} \partial_{k_a}\varepsilon_n}{\omega_1} + \sum_{nm} f_{nm}\varepsilon_{mn}\frac{\A^b_{nm}\A^a_{mn}}{\omega_1+\omega_{mn}}
    \bigg\}
\end{align*}
where $\hbar\omega_{mn} = \varepsilon_m - \varepsilon_n$. The two terms can be understood as a Drude part, which appears only if there is a Fermi surface, and an interband part, depending on $Q^{mn}_{ba} = \A^b_{nm}\A^a_{mn}$. The second-order conductivity is likewise given by
\begin{align*}
    \sigma^{a;bc}(\omega_1,\omega_2) &= \frac{e^3}{\hbar^3}\int_k 
    \bigg\{
        \frac{1}{\omega_1 \omega_2} \sum_n f_n \partial_{k_c}\partial_{k_b} \partial_{k_a} \varepsilon_n
        \\
        &-\sum_{nm} f_{nm}
        \left[
            \partial_{k_c} \frac{\varepsilon_{nm}\A_{nm}^a\A_{mn}^b}{\omega_1(\omega_2+\omega_{nm})}
            - \frac{\A_{nm}^c}{\omega_1+\omega_{mn}} \partial_{k_b} \frac{\varepsilon_{mn}\A_{mn}^a}{\omega_2+\omega_{mn}}
        \right]
        \\
        &-\sum_{nml} f_{nm}
        \left[
            \frac{v_{nl}^a\A_{lm}^b\A_{mn}^c}{(\omega_1+\omega_{nm})(\omega_2+\omega_{nl})}  + \frac{\A_{nm}^c\A_{ml}^bv_{ln}^a}{(\omega_1+\omega_{mn})(\omega_2+\omega_{ln})}
        \right]
    \bigg\}
\end{align*}
This expression agrees with the literature~\cite{Passos_2021,aversaNonlinearOpticalSusceptibilities}, however, often a different frequency convention is used, as clarified in the following section.

\section{Optical conductivity frequency conventions}
In our work, the frequencies $\omega_1$ and $\omega_2$ arise from Fourier transforming $\tau_1$ and $\tau_2$, which correspond to time differences, and are the natural choice of time arguments in the context of 2DCS. In the non-2DCS literature, the conductivity is often written in terms of different frequencies $\omega_A$ and $\omega_B$, which correspond to the alternative time arguments $t_A=\tau_1 + \tau_2$ and $t_B=\tau_2$. These different conventions are related by the mapping $\omega_1 = \omega_A$, $\omega_2 = \omega_A + \omega_B$. This can be seen by comparing definitions: As written above and in the main text, we define the second order optical conductivity in the time domain according to:
\begin{align*}
    j_2^{a}(t) &= \sum_{bc} \int d\tau_1 d\tau_2 \sigma^{a;bc}(\tau_1,\tau_2)E^b(t-\tau_2)E^c(t-\tau_1-\tau_2) 
    .
\end{align*}
The frequency dependent conductivity is then given by the Fourier transform:
\begin{align*}
    \sigma^{a;bc}(\omega_1,\omega_2) = \int d\tau_1 d\tau_2 e^{i\omega_1\tau_1}e^{i\omega_2\tau_2}\sigma^{a;bc}(\tau_1,\tau_2)
    .
\end{align*}
In the alternative convention, the time-domain conductivity is defined according to
\begin{align}
    \label{eq:tdcond_alt}
    j_2^{a}(t) &= \sum_{bc} \int dt_A dt_B \tilde{\sigma}^{a;bc}(t_A,t_B)E^b(t-t_B)E^c(t-t_A) 
    ,
\end{align}
where $\tilde{\sigma}$ is the conductivity written in the alternative convention. The corresponding frequency-dependent conductivity is similarly defined as:
\begin{align*}
    \tilde{\sigma}^{a;bc}(\omega_A,\omega_B) = \int dt_A dt_B e^{i\omega_At_A}e^{i\omega_Bt_B}\tilde{\sigma}^{a;bc}(t_A,t_B)
    .
\end{align*}
To compare conventions, we note that we can use the Fourier transform to rewrite (\ref{eq:tdcond_alt}) as
\begin{align*}
    j_2^{a}(t) &= \sum_{bc} \int dt_A dt_B \tilde{\sigma}^{a;bc}(t_A,t_B)E^b(t-t_B)E^c(t-t_A)  \\
    &=\sum_{bc} \int dt_A dt_B \tilde{\sigma}^{a;bc}(t_A,t_B) \left[\int \frac{d\omega_B}{2\pi}e^{-i\omega_B(t-t_B)}E^b(\omega_B)\right]\left[\int \frac{d\omega_A}{2\pi}e^{-i\omega_A(t-t_A)}E^c(\omega_A)\right]\\
    &= \sum_{bc}\int \frac{d\omega_A}{2\pi}\frac{d\omega_B}{2\pi} \tilde{\sigma}^{a;bc}(\omega_A,\omega_B) E^b(\omega_B)E^c(\omega_A)e^{-i(\omega_A + \omega_B)t}
\end{align*}
Similarly, we find:
\begin{align*}
    j_2^{a}(t) &= \sum_{bc} \int d\tau_1 d\tau_2 \sigma^{a;bc}(\tau_1,\tau_2)E^b(t-\tau_2)E^c(t-\tau_1-\tau_2) \\
    &= \sum_{bc} \int d\tau_1 d\tau_2 \sigma^{a;bc}(\tau_1,\tau_2)\left[\int \frac{d\omega_B}{2\pi} e^{-i\omega_B(t-\tau_2)} E^b(\omega_B)\right] \left[\int \frac{d\omega_A}{2\pi} e^{-i\omega_A(t-\tau_1-\tau_2)} E^c(\omega_A)\right] \\
    &= \sum_{bc}\int \frac{d\omega_A}{2\pi} \frac{d\omega_B}{2\pi} \sigma^{a;bc}(\omega_A,\omega_A + \omega_B) E^b(\omega_B)E^c(\omega_A)e^{-i(\omega_A + \omega_B)t}
\end{align*}
from which we can read off $\omega_1 = \omega_A, \omega_2 = \omega_A + \omega_B$.

The frequencies in the above expressions should all be understood to contain small imaginary contributions $\omega_1 \to \omega_1 + i\eta, \omega_2 \to \omega_2 + 2i\eta$. Physically, $\eta$ can be understood as a phenomenological scattering rate, and can be formally incorporated by turning on the perturbing field adiabatically $E^a(t) \to e^{i\eta t}E^a(t)$, or by introducing a scattering term into the density matrix equation of motion. The factor of $2\eta$ for $\omega_2$ is consistent with $\omega_A \to \omega_A + i \eta$, $\omega_B \to \omega_B + i \eta$ in the other frequency convention.

\section{Geometric decomposition }
The geometric decomposition given in the main text is recovered after the following manipulations. First, we integrate the first three terms by parts, making the Fermi surface contributions explicit. The expression becomes
\begin{align*}
    \sigma^{a;bc}(\omega_1,\omega_2) = \frac{e^3}{\hbar^3}\int_k 
    \bigg\{
        &-g_0^{\omega_1}g_0^{\omega_2} \sum_n (\partial_{k_c}f_n) (\partial_{k_b} \partial_{k_a} \varepsilon_n)
        + g_0^{\omega_1}\sum_{nm} g_{nm}^{\omega_2}(\partial_{k_c} f_{nm})\varepsilon_{nm}\A^a_{nm}\A^b_{mn}
        \\
        &
        + \sum_{nm}g_{mn}^{\omega_1}g_{mn}^{\omega_2}(\partial_{k_b} f_{nm})\varepsilon_{nm}\A_{nm}^c\A_{mn}^a 
        +\sum_{nm} (\partial_{k_b} g_{mn}^{\omega_1})g_{mn}^{\omega_2}f_{nm}\varepsilon_{nm} \A_{nm}^c\A_{mn}^a \\
        &
        + \sum_{nm} g_{mn}^{\omega_1}g_{mn}^{\omega_2}f_{nm}\varepsilon_{nm}(\partial_{k_b} \A_{nm}^c)\A_{mn}^a
        \\
        &-\hbar \sum_{nml} f_{nm}
        \left[
            g_{nm}^{\omega_1}g_{nl}^{\omega_2}
            v_{nl}^a\A_{lm}^b\A_{mn}^c  + g_{mn}^{\omega_1}g_{ln}^{\omega_2}\A_{nm}^c\A_{ml}^bv_{ln}^a
        \right]
    \bigg\},
\end{align*}
where $g_0^\omega = 1/\omega$ and  $g_{mn}^\omega = 1/(\omega + \omega_{mn}).$ Next, we rewrite the last term by relabelling $(n \leftrightarrow m)$ in the first term, expanding out the velocity matrix element $v_{nm}^a = \frac{1}{\hbar} \left( \delta_{nm}\partial_{k_a}\varepsilon_n + i \varepsilon_{nm}\A^a_{nm}\right)$, and pulling the $l=n$ and $l=m$ terms out of the sum: 
\begin{align*}
&-\hbar\sum_{nml} f_{nm}
        \left[
            g_{nm}^{\omega_1}g_{nl}^{\omega_2}
            v_{nl}^a\A_{lm}^b\A_{mn}^c  + g_{mn}^{\omega_1}g_{ln}^{\omega_2}\A_{nm}^c\A_{ml}^bv_{ln}^a
        \right]\\
        &=
   \hbar\sum_{nml} g_{mn}^{\omega_1}f_{nm}\A_{nm}^c
        \left[
            g_{ml}^{\omega_2}
            v_{ml}^a\A_{ln}^b  -g_{ln}^{\omega_2}\A_{ml}^bv_{ln}^a
        \right] \\
        &=
        \hbar\sum_{nm}
        \bigg\{
        g_{mn}^{\omega_1}f_{nm}\A_{nm}^c
        \left[
            g_{mn}^{\omega_2}
            v_{mn}^a\A_{nn}^b  -g_{0}^{\omega_2}\A_{mn}^bv_{nn}^a
            +
            g_{0}^{\omega_2}
            v_{mm}^a\A_{mn}^b  -g_{mn}^{\omega_2}\A_{mm}^bv_{mn}^a
        \right]
        \\
        &+
      \hbar\sum_{l\neq n,m} g_{mn}^{\omega_1}f_{nm}\A_{nm}^c
        \left[
            g_{ml}^{\omega_2}
            v_{ml}^a\A_{ln}^b  -g_{ln}^{\omega_2}\A_{ml}^bv_{ln}^a
        \right] 
        \bigg\}
    \\
    &=
    \sum_{nm}
        \bigg\{
        g_{mn}^{\omega_1}f_{nm}\A_{nm}^c
        \left[
            ig_{mn}^{\omega_2}
            \varepsilon_{nm}\A_{mn}^a(\A_{mm}^b - \A_{nn}^b)   
            + g_{0}^{\omega_2}
            (\partial_{k_a}\varepsilon_{mn})\A_{mn}^b 
        \right]
        \\
        &+
      i\sum_{l\neq n,m} g_{mn}^{\omega_1}f_{nm}\A_{nm}^c
        \left[
            g_{ml}^{\omega_2}
            \varepsilon_{ml}\A_{ml}^a\A_{ln}^b  -g_{ln}^{\omega_2}\varepsilon_{ln}\A_{ml}^b\A_{ln}^a
        \right] 
        \bigg\}
\end{align*}
Reinserting this into the full expression, we find:
\begin{align*}
    \sigma^{a;bc}(\omega_1,\omega_2) = \frac{e^3}{\hbar^3}\int_k 
    \bigg\{
        &-g_0^{\omega_1}g_0^{\omega_2} \sum_n (\partial_{k_c}f_n) (\partial_{k_b} \partial_{k_a} \varepsilon_n)
        + g_0^{\omega_1}\sum_{nm} g_{nm}^{\omega_2}(\partial_{k_c} f_{nm})\varepsilon_{nm}\A^a_{nm}\A^b_{mn}
        \\
        &
        + \sum_{nm}g_{mn}^{\omega_1}g_{mn}^{\omega_2}(\partial_{k_b} f_{nm})\varepsilon_{nm}\A_{nm}^c\A_{mn}^a 
        +\sum_{nm} (\partial_{k_b} g_{mn}^{\omega_1})g_{mn}^{\omega_2}f_{nm}\varepsilon_{nm} \A_{nm}^c\A_{mn}^a \\
        &
        + \sum_{nm} g_{mn}^{\omega_1}g_{mn}^{\omega_2}f_{nm}\varepsilon_{nm}\A_{mn}^a(\partial_{k_b} \A_{nm}^c - i(\A^b_{nn} - \A^b_{mm})\A^c_{nm})
        \\
        &+ g_0^{\omega_2}\sum_{nm}g_{mn}^{\omega_1}f_{nm}(\partial_{k_a} \varepsilon_{mn})\A_{nm}^c\A_{mn}^b
        \\
        &
        +
      i\sum_{nm} g_{mn}^{\omega_1}f_{nm}\A_{nm}^c
      \sum_{l\neq n,m}
        \left[
            g_{ml}^{\omega_2}
            \varepsilon_{ml}\A_{ml}^a\A_{ln}^b  -g_{ln}^{\omega_2}\varepsilon_{ln}\A_{ml}^b\A_{ln}^a
        \right] 
        \bigg\}
\end{align*}
Finally, recalling the definitions $Q_{ab}^{mn} = \A^a_{nm}\A^b_{mn}$ and $C_{a;bc}^{mn} = \A_{nm}^b[\partial_{k_a}\A^c_{mn} - i(\A^a_{mm}-\A^a_{nn})\A_{mn}^c]$, we recover the decomposition given in the main text:
\begin{align*}
    \sigma^{a;bc}(\omega_1,\omega_2) = \frac{e^3}{\hbar^3}\int_k 
    \bigg\{
        &-g_0^{\omega_1}g_0^{\omega_2} \sum_n (\partial_{k_c}f_n) (\partial_{k_b} \partial_{k_a} \varepsilon_n)
        + g_0^{\omega_1}\sum_{nm} g_{nm}^{\omega_2}(\partial_{k_c} f_{nm})\varepsilon_{nm}Q_{ab}^{mn}
        \\
        &
        + \sum_{nm}g_{mn}^{\omega_1}g_{mn}^{\omega_2}(\partial_{k_b} f_{nm})\varepsilon_{nm}Q_{ca}^{mn}
        +\sum_{nm} (\partial_{k_b} g_{mn}^{\omega_1})g_{mn}^{\omega_2}f_{nm}\varepsilon_{nm}Q_{ca}^{mn} \\
        &
        + \sum_{nm} g_{mn}^{\omega_1}g_{mn}^{\omega_2}f_{nm}\varepsilon_{nm}C^{nm}_{b;ac}
        \\
        &+ g_0^{\omega_2}\sum_{nm}g_{mn}^{\omega_1}f_{nm}(\partial_{k_a} \varepsilon_{mn})Q_{cb}^{mn}
        \\
        &
        +
      i\sum_{nm} g_{mn}^{\omega_1}f_{nm}\A_{nm}^c
      \sum_{l\neq n,m}
        \left[
            g_{ml}^{\omega_2}
            \varepsilon_{ml}\A_{ml}^a\A_{ln}^b  -g_{ln}^{\omega_2}\varepsilon_{ln}\A_{ml}^b\A_{ln}^a
        \right] 
        \bigg\}
\end{align*}

\section{Isolating the quantum connection term}
In time reversal symmetric systems, $\varepsilon_n(\vk) = \varepsilon_n(-\vk)$ and $\A^a_{nm}(\vk) = \A_{mn}^a(-\vk)$. It follows that $f_n(\vk) = f_n(-\vk), Q^{mn}_{ab}(\vk) = [Q^{mn}_{ab}(-\vk)]^*$, and $C_{a;bc}^{mn}(\vk) = -[C_{a;bc}^{mn}(-\vk)]^*$. We next note that $Q^{mn}_{aa} = \A_{nm}^a\A_{mn}^a = |\A_{mn}^a|^2$ is purely real, and is thus even under $\vk \to -\vk$. Under these conditions it is clear that $\sigma^{a;aa}_\mathrm{Drude}, \sigma^{a;aa}_\mathrm{an},\sigma^{a;aa}_\mathrm{DR},\sigma^{a;aa}_\mathrm{HOP},$ and $\sigma^{a;aa}_\mathrm{inj}$ are integrals of functions odd in momentum and therefore vanish. 

To distinguish the $\sigma_C$ term from the $\sigma_\mathrm{3B}$ term, we note that that $\sigma_C$ is peaked along the diagonal, while $\sigma_\mathrm{3B}$ is peaked at frequencies satisfying 
\begin{equation}
\label{eq:c1}
(\hbar\omega_1, \hbar\omega_2) =(\varepsilon_{mn},\varepsilon_{ml})
\end{equation}
or 
\begin{equation}
\label{eq:c2}
    (\hbar\omega_1, \hbar\omega_2) = (\varepsilon_{mn},\varepsilon_{ln})
    .
\end{equation} These conditions coincide with the frequency diagonal only when $\varepsilon_l = \varepsilon_n$ or $\varepsilon_l = \varepsilon_m$. However, the 3B term vanishes for degenerate bands. Explicitly, given $\varepsilon_l = \varepsilon_n$, the first part of the 3B term in the $a=b=c$ limit is of the form
\begin{align*}
    \frac{ie^3}{\hbar^3}\int_k\sum_{nm} g_{mn}^{\omega_1}f_{nm}\A^a_{nm}g_{ml}^{\omega_2}\varepsilon_{ml}\A^a_{ml}\A^a_{ln}
    &\to
    \frac{ie^3}{\hbar^3}\int_k\sum_{nm} g_{mn}^{\omega_1}f_{nm}\A^a_{nm}g_{mn}^{\omega_2}\varepsilon_{mn}\A^a_{ml}\A^a_{ln} \\
    &= 
    -\frac{ie^3}{\hbar^3}\int_k\sum_{nm} g_{mn}^{-\omega_1}f_{nm}(\A^a_{nm})^*g_{mn}^{-\omega_2}\varepsilon_{mn}(\A^a_{ml}\A^a_{ln})^* \\
    &=
    -\frac{ie^3}{\hbar^3}\int_k\sum_{nm} g_{nm}^{\omega_1}f_{nm}\A^a_{mn}g_{nm}^{\omega_2}\varepsilon_{mn}\A^a_{lm}\A^a_{nl} \\
    &=
    -\frac{ie^3}{\hbar^3}\int_k\sum_{nm} g_{mn}^{\omega_1}f_{nm}\A^a_{nm}g_{mn}^{\omega_2}\varepsilon_{mn}\A^a_{ml}\A^a_{ln}
\end{align*}
where we have used the fact that $\sigma(\omega_1,\omega_2) = \sigma^*(-\omega_1,-\omega_2)$. A similar argument holds for the second term in the case where $\varepsilon_l = \varepsilon_m$. Thus, we have shown that $\sigma_\mathrm{3B}$ vanishes along the diagonal. Furthermore, in our examples, the conditions (\ref{eq:c1}) and (\ref{eq:c2}) ensure the 3B term is peaked only far from the diagonal and at large frequencies, and so will be numerically small in the vicinity of the diagonal. Therefore we can unambiguously identify the 2DCS signal $\sigma^{a;aa}(\omega, \omega)$ as arising from the quantum connection.

\section{Projector formalism}
The non-Abelian Berry connection $\A^a_{nm}$ is not invariant under the gauge transformation $|\psi_{n\vk}\rangle \to e^{i\theta_{n\vk}}|\psi_{n\vk}\rangle$, making numerical calculations inconvenient. However, geometric quantities like the quantum geometric tensor and quantum connection can be expressed in a gauge-invariant way using the projector formalism~\cite{avdoshkinMultistateGeometryShift2024,mitscherlingGaugeinvariantProjectorCalculus2024}. We introduce the projector onto the $n$-th band,
\begin{equation}
    P_n(\vk) = |u_{n\vk}\rangle \langle u_{n\vk}|
    ,
\end{equation}
in terms of which the multi-band QGT and quantum connection can be written as follows:
\begin{equation}
    Q_{ab}^{mn} = -\tr\left(P_n (\partial_{k_a} P_m)(\partial_{k_b} P_n)\right)
\end{equation}
\begin{equation}
    C_{a;bc}^{mn} = -\tr \left(P_n(\partial_{k_b} P_m)
    [(\partial_{k_a}\partial_{k_c}P_n) + (\partial_{k_a} P_m)(\partial_{k_c} P_n)]\right)
    ,
\end{equation}
where here the trace is over the cell-periodic states at a single $\vk$ point: $\tr(\cdot) = \sum_n \langle u_{n\vk}| \cdot |u_{n\vk}\rangle$
The triple products of Berry connections appearing in the $\sigma_\mathrm{3B}$ term can also be written using projectors via
\begin{equation}
    \A^a_{nm}\A^b_{ml}\A^c_{ln} = -i\tr(P_n (\partial_{k_a} P_m)(\partial_{k_b} P_l)(\partial_{k_c} P_n))
\end{equation}
where $m\neq n \neq l$. We have adopted the sign convention of~\cite{ahnRiemannianGeometryResonant2022}, which differs by a minus sign when compared with~\cite{avdoshkinMultistateGeometryShift2024}. We employ the projector formalism throughout this work for performing the numerical calculations.

\section{Transition metal dichalcogenide model}
Here we provide details on the three band tight-binding TMD models used as examples in our work. We include the Hamiltonians and parameters here for reference, and refer the reader to the original source for further details~\cite{liuThreebandTightbindingModel2013}.

These models are constructed by considering the symmetry-allowed hopping terms between the relevant $d_{z^2}, d_{xy}$, and $d_{x^2-y^2}$ orbitals of the transition metal ions. If only nearest-neighbour (NN) hoppings are considered, the Hamiltonian takes the form
    \begin{equation*}
        H_\mathrm{TMD}^\mathrm{NN}(\vk) = 
        \begin{bmatrix}
            h_0 & h_1 & h_2 \\
            h_1^* & h_{11} & h_{12} \\
            h_2^* & h_{12}* & h_{22}
        \end{bmatrix}
    \end{equation*}
with the matrix elements given by
    \begin{align*}
    h_0 &= 2t_0(\cos2\alpha + 2\cos\alpha\cos\beta) + \epsilon_1,\\
    h_1 &= -2 \sqrt{3} t_2 \sin\alpha \sin \beta + 2it_1(\sin 2\alpha + \sin\alpha \cos\beta), \\
    h_2 &= 2t_2(\cos2\alpha - \cos\alpha\cos\beta) + 2\sqrt{3}it_1 \cos\alpha \sin\beta, \\
    h_{11} &= 2t_{11} \cos 2\alpha + (t_{11} + 3t_{22}) \cos\alpha \cos\beta + \epsilon_2, \\
    h_{22} &= 2t_{22} \cos 2\alpha + (3t_{11} + t_{22}) \cos\alpha \cos\beta + \epsilon_2, \\
    h_{12} &= \sqrt{3}(t_{22} - t_{11}) \sin\alpha \sin\beta + 4it_{12} \sin\alpha(\cos\alpha - \cos\beta),
    \end{align*}
where we have defined
    \begin{align*}
        (\alpha,\beta) = \left(\frac{k_x}{2},\frac{\sqrt{3}k_y}{2}\right)
        .
    \end{align*}
    
The parameters of the model are determined by fitting the bands to first principles calculations. For the first example in our work, we use the parameters given in Table~\ref{tab:TMD}, which result from fitting to the MoTe$_2$ generalized-gradient approximation (GGA) band structure.

The accuracy of the three band model is improved by considering up to third-nearest neighbour (TNN) hoppings. This is the case considered in our second example, where we compare MoTe$_2$ GGA, WTe$_2$ GGA, and MoTe$_2$ LDA. The Hamiltonian is given by
    \begin{equation*}
        H_\mathrm{TMD}^\mathrm{TNN}(\vk) = 
        \begin{bmatrix}
            V_0 & V_1 & V_2 \\
            V_1^* & V_{11} & V_{12} \\
            V_2^* & V_{12}* & V_{22}
        \end{bmatrix}
        ,
    \end{equation*}
with matrix elements
    \begin{align*}
        V_0 &= \epsilon_1 + 2t_0(2\cos\alpha \cos\beta + \cos2\alpha) + 2r_0(2\cos3\alpha \cos\beta + \cos2\beta) + 2u_0(2\cos2\alpha \cos2\beta + \cos4\alpha),\\
        \Re V_1 &= -2\sqrt{3}t_2 \sin\alpha \sin\beta + 2(r_1 + r_2) \sin3\alpha \sin\beta - 2\sqrt{3}u_2 \sin2\alpha \sin2\beta,\\
        \Im V_1 &= 2t_1 \sin \alpha(2 \cos\alpha + \cos\beta) + 2(r_1 - r_2)\sin3\alpha \cos\beta + 2u_1\sin2\alpha(2\cos2\alpha + \cos2\beta), \\
        \Re V_2 &= 2t_2(\cos2\alpha - \cos\alpha\cos\beta) - \frac{2}{\sqrt{3}} (r1 + r2)(\cos3\alpha\cos\beta - \cos2\beta) + 2u_2(\cos4\alpha - \cos2\alpha \cos2\beta),\\
        \Im V_2 &= 2\sqrt{3}t_1 \cos\alpha \sin\beta + \frac{2}{\sqrt{3}} \sin\beta(r_1 - r_2)(\cos3\alpha + 2\cos\beta) + 2\sqrt{3}u_1\cos2\alpha \sin2\beta,\\
        V_{11} &= \epsilon_2 + (t_{11} + 3t_{22})\cos\alpha \cos\beta + 2t_{11} \cos2\alpha + 4r_{11}\cos3\alpha \cos\beta + 2(r_{11} + \sqrt{3}r_{12}) \cos2\beta \nonumber, \\
        &+ (u_{11} + 3u_{22}) \cos2\alpha \cos2\beta + 2u_{11} \cos4\alpha\\
        \Re V_{12} &= \sqrt{3}(t_{22} - t_{11})\sin\alpha \sin\beta + 4r_{12}\sin3\alpha \sin\beta + \sqrt{3}(u_{22} - u_{11})\sin2\alpha \sin2\beta,\\
        \Im V_{12} &= 4t_{12}\sin\alpha(\cos\alpha - \cos\beta) + 4u_{12} \sin2\alpha(\cos2\alpha - \cos2\beta),\\
            V_{22} &= \epsilon_2 + (3t_{11} + t_{22})\cos\alpha \cos\beta + 2t_{22}\cos2\alpha + 2r_{11}(2\cos3\alpha\cos\beta + \cos2\beta) \nonumber, \\
            &+ \frac{2}{\sqrt{3}}r_{12}(4\cos3\alpha \cos\beta - \cos2\beta) + (3u_{11} + u_{22})\cos2\alpha \cos2\beta + 2u_{22}\cos4\alpha
            .
    \end{align*}
The relevant fitted parameters are listed in Table~\ref{tab:TMD}.
    \begin{table}[h!]
    \centering
    \begin{tabular*}{0.8\textwidth}{@{\extracolsep{\fill}}c|cccccccccc}
         & $\epsilon_1$ & $\epsilon_2$ & $t_0$ & $t_1$ & $t_2$ & $t_{11}$ & $t_{12}$ & $t_{22}$ & $r_0$ & $r_1$ \\
         & $r_2$ & $r_{11}$ & $r_{12}$ & $u_0$ & $u_1$ & $u_2$ & $u_{11}$ & $u_{12}$ & $u_{22}$ \\
        \hline
        MoTe$_2$ NN GGA & 0.605 & 1.972 & -0.169 & 0.228 & 0.390 & 0.207 & 0.239 & 0.252 & - & - \\
        & - & - & - & - & - & - & - & - & - \\
        MoTe$_2$ TNN GGA & 0.588 & 1.303 & -0.226 & -0.234 & 0.036 & 0.400 & 0.098 & 0.017 & 0.003 & -0.025 \\
        & -0.169 & 0.082 & 0.051 & 0.057 & 0.103 & 0.187 & -0.045 & -0.141 & 0.087\\
        MoTe$_2$ TNN LDA & 0.574 & 1.410 & -0.148 & -0.173 & 0.333 & 0.203 & 0.186 & 0.127 &  0.007 & -0.280 \\
        & 0.067 &  0.073 &  0.081 & -0.054 & 0.008 &  0.037 &  0.145 & -0.078 &  0.035 \\
        WTe$_2$ TNN GGA & 0.697 & 1.380 & -0.109 & -0.164 & 0.368 & 0.204 & 0.093 & 0.038 & -0.015 & -0.209 \\
        & 0.107 & 0.115 & 0.009 & -0.066 & 0.011 & -0.013 & 0.312 & -0.177 & -0.132
    \end{tabular*}
    \caption{TMD parameters (in eV) for both NN and TNN tight binding models \cite{liuThreebandTightbindingModel2013}.}
    \label{tab:TMD}
    \end{table}
    
\section{Sr$_2$RuO$_4$ model}
Finally, we provide the model and further details regarding the Sr$_2$RuO$_4$ model introduced in the main text. The three band model is derived by considering the $t_{2g}$ $d$ orbitals of Ru ions on a square lattice, considering both kinetic and spin-orbit coupling terms~\cite{puetterIdentifyingSpintripletPairing2012}. In the $(c^{yz\dagger}_{\vk\sigma}, c^{xz\dagger}_{vk\sigma}, c^{xy\dagger}_{\vk,-\sigma})$ basis, the Hamiltonian is given by
    \begin{equation*}
        H_\mathrm{SR}(\vk, \sigma) = 
        \begin{bmatrix}
            \varepsilon_\vk^{yz} & \varepsilon_\vk^\mathrm{1d} + i\sigma \lambda & -\sigma \lambda \\
            \varepsilon_\vk^\mathrm{1d} - i \sigma \lambda & \varepsilon_\vk^{xz} & i \lambda\\
             -\sigma \lambda & - i\lambda & \varepsilon_\vk^{xy}
        \end{bmatrix}
        ,
    \end{equation*}
with dispersions
    \begin{gather*}
        \varepsilon_\vk^{yz/xz} = -2t_1\cos k_{y/x} - 2t_2\cos k_{x/y} - \mu_1 \\
        \varepsilon_\vk^{xy} = -2t_3(\cos k_x + \cos k_y ) - 4t_4\cos k_x \cos k_y - \mu_2 \\
        \varepsilon_\vk^\mathrm{1d} = -4t_5\sin k_x \sin k_y
        .
    \end{gather*}
This Hamiltonian is inversion symmetric, and thus the second-order optical response vanishes. In order to break inversion symmetry in the model, we introduce the following perturbation: 
    \begin{equation*}
        H_\mathrm{SR}'(\vk) = 
        \begin{bmatrix}
            0 & 2it_5\delta e^{ik_x}\sin k_y &0 \\
            -2it_5\delta e^{-ik_x}\sin k_y & 0 & 0\\
             0 & 0& 0
        \end{bmatrix}
    \end{equation*}
After adding this perturbation, the bands undergo a small spin splitting, and thus the three band model is really a six band model. As a result, we must add spin labels to all the relevant quantities: $\varepsilon_n \to \varepsilon_{n\sigma}, \A^a_{nm} \to \A^a_{nm,\sigma\sigma'} := i\langle u_{n\sigma}|\partial_{k_a} u_{m\sigma'}\rangle$. However, since the Hamiltonian is block diagonal in $\sigma$, the non-Abelian Berry connection vanishes if $\sigma \neq \sigma'$. Therefore to correct the expression for the relevant conductivity (c.f. (\ref{eq:TRS_sigma}) in the main text), we simply need to add a single spin label:
\begin{equation*}
    \sigma^{a;aa}(\omega_1,\omega_2) = ie^3 \int_k \sum_{nm,\sigma} g_{mn,\sigma}^{\omega_1}g_{mn,\sigma}^{\omega_2}f_{nm,\sigma}\tilde{\Gamma}_{a;aa}^{nm,\sigma}
\end{equation*}
where $g_{mn,\sigma}^{\omega}$ and $f_{mn,\sigma}$ depend on $\varepsilon_{mn,\sigma} = \varepsilon_{m\sigma} - \varepsilon_{n\sigma}$ and $\tilde{\Gamma}_{a;aa}^{nm,\sigma} = \operatorname{Im} \A^a_{mn,\sigma}(\partial_{k_a}\A_{nm,\sigma}^a - i(\A_{nn,\sigma}^a - \A^a_{mm,\sigma})\A_{nm,\sigma}^a)$. Including spin, time-reversal symmetry enforces $\varepsilon_{n,\sigma}(\vk) = \varepsilon_{n,-\sigma}(-\vk)$ and $\A^a_{nm,\sigma}(\vk) = \A^a_{mn,-\sigma}(-\vk)$, which in turn enforce $f_{n\sigma}(\vk) = f_{n,-\sigma}(-\vk), g_{mn,\sigma}^{\omega}(\vk) = g_{mn,-\sigma}^{\omega}(-\vk)$ and $\tilde{\Gamma}_{a;aa}^{nm,\sigma}(\vk) = \tilde{\Gamma}_{a;aa}^{nm,-\sigma}(-\vk)$. Hence we can rewrite the conductivity as
\begin{align*}
    \sigma^{a;aa}(\omega_1,\omega_2) &= ie^3 \int_k \sum_{nm} g_{mn,\uparrow}^{\omega_1}(\vk)g_{mn,\uparrow}^{\omega_2}(\vk)f_{nm,\uparrow}(\vk)\tilde{\Gamma}_{a;aa}^{nm,\uparrow}(\vk) + g_{mn,\downarrow}^{\omega_1}(\vk)g_{mn,\downarrow}^{\omega_2}(\vk)f_{nm,\downarrow}(\vk)\tilde{\Gamma}_{a;aa}^{nm,\downarrow}(\vk) \\
    &= ie^3 \int_k \sum_{nm} g_{mn,\uparrow}^{\omega_1}(\vk)g_{mn,\uparrow}^{\omega_2}(\vk)f_{nm,\uparrow}(\vk)\tilde{\Gamma}_{a;aa}^{nm,\uparrow}(\vk) + g_{mn,\uparrow}^{\omega_1}(-\vk)g_{mn,\uparrow}^{\omega_2}(-\vk)f_{nm,\uparrow}(-\vk)\tilde{\Gamma}_{a;aa}^{nm,\uparrow}(-\vk) \\
    &= 2ie^3 \int_k \sum_{nm} g_{mn,\uparrow}^{\omega_1}(\vk)g_{mn,\uparrow}^{\omega_2}(\vk)f_{nm,\uparrow}(\vk)\tilde{\Gamma}_{a;aa}^{nm,\uparrow}(\vk)
\end{align*}
Therefore it is clear we can work exclusively with the three $\uparrow$ bands, which is what is done in the main text, where the spin label has been suppressed, and the extra factor of 2 has been included in calculations. The parameters used for this model are given below in Table~\ref{tab:sr2ruo4}.
    \begin{table}[h!]
    \centering
    \begin{tabular*}{0.8\textwidth}{@{\extracolsep{\fill}}ccccccccc}
         $t_1$ & $t_2$ & $t_3$ & $t_4$ & $t_5$ & $\lambda$ & $\mu_1$ & $\mu_2$ & $\delta$ \\
        \hline
        0.252 & 0.028 & 0.280 & 0.112 & 0.014 & 0.048 & 0.297 & 0.353 & 0.3
    \end{tabular*}
    \caption{Sr$_2$RuO$_4$ parameters adapted from~\cite{Lindquist_2020} (in eV, except $\delta$ which is unitless)}
    \label{tab:sr2ruo4}
    \end{table}

\end{appendix}
\end{document}